\documentclass[aps,prd,onecolumn,groupedaddress,showpacs,showkeys, nofootinbib,amssymb]{revtex4}
\usepackage[T1]{fontenc}
\usepackage[latin1]{inputenc}
\usepackage{graphicx}
\usepackage[english]{babel}
\usepackage{amsmath}
\usepackage{amssymb}
\usepackage{amsfonts}

\begin{document}
\def\pp{{\, \mid \hskip -1.5mm =}}
\def\cL{{\cal L}} 
\def\beq{\begin{equation}}
\def\eneq{\end{equation}}
\def\bea{\begin{eqnarray}}
\def\enea{\end{eqnarray}}
\def\tr{{\rm tr}\, }
\def\nn{\nonumber \\}
\def\e{{\rm e}}

\title{\textbf{Equivalence between Palatini and metric formalisms of  $f(R)$-gravity by divergence free current}}

\author{Salvatore Capozziello$^{1,2}$, Farhad Darabi$^3$, Daniele Vernieri$^1$}

\affiliation{\it $^1$Dipartimento di Scienze Fisiche, Università
di Napoli {}``Federico II'',  $^2$INFN Sez. di Napoli, Compl. Univ. di
Monte S. Angelo, Edificio G, Via Cinthia, I-80126, Napoli, Italy \\ 
$^3$Department of Physics, Azarbaijan University of Tarbiat Moallem, Tabriz 53741-161, Iran\\
Research Institute for Astronomy and Astrophysics of Maragha (RIAAM), Maragha 55134-441, Iran}

\date{\today}

\begin{abstract}
The equivalence between metric and Palatini formalisms in $f(R)$-gravity  can be achieved in the 
general context of theories with divergence free current. This equivalence is a necessary result of a symmetry which is included 
in a particular conservation equation of the current. In fact the conservation equation, by an appropriate 
redefinition of the introduced auxiliary field, may be encoded in a massless scalar field equation.
\end{abstract}

\maketitle

\section{introduction}
\label{1}

Extended Theories of Gravity  (ETGs) \cite{mauro, odintsov, farasot, libroSaFe,defelice}  address several shortcomings of standard General Relativity (GR) \cite{wald, wein, misn} related to cosmology, astrophysics and quantum field theory. The idea to extend Einstein's theory of gravitation is fruitful and economic with respect to several attempts which try to solve problems by adding new and often unjustified ingredients in order to give a self-consistent picture of  dynamics. The present-day observed accelerated expansion of Hubble flow and the missing matter of astrophysical structures are primarily enclosed in these considerations \cite{JCAP,mnras}. The general paradigm to solve both issues consists in adding, into the effective action, physically
motivated higher order curvature invariants and non-minimally coupled scalar fields \cite{odi,farh}.
An alternative philosophy is to add new cosmic fluids (new components in the rhs of the field equations) which should give rise to clustered structures (dark matter) or to accelerated dynamics (dark energy) thanks to exotic equations of state. In particular, relaxing the hypothesis that gravitational Lagrangian has to be a linear function of the Ricci curvature scalar $R$, as in the Hilbert-Einstein formulation, one can take into account, as a minimal extension, an effective action where the gravitational Lagrangian is a generic $f(R)$ function.

Furthermore,  one can consider more general actions in which we have a scalar field non-minimally coupled to the gravity, considering them as a generalization of Brans-Dicke theory \cite{cimall, brans}. By conformal transformations,  it is possible to show that any higher-order or scalar-tensor theory, in absence of ordinary matter, e.g.. a perfect fluid, is conformally equivalent to the Einstein theory plus minimally coupled scalar fields. The converse is also true in fact we can transform a standard Einstein theory into a non-minimally coupled scalar-tensor theory. The problem should be faced from a more general viewpoint and the Palatini approach to gravity could be useful to this goal. The fundamental idea of the Palatini formalism is to consider the (usually torsionless)  connection $\Gamma$, entering the definition of the Ricci tensor, to be independent of the metric $g$ defined on the spacetime $\mathcal{M}$. The Palatini formulation for the standard Hilbert-Einstein theory results to be equivalent to the purely metric theory: this follows from the fact that the field equations for the connection $\Gamma$, firstly considered to be independent of the metric, give exactly the Levi-Civita connection of the metric $g$ in the Hilbert-Einstein case. As a consequence, there is no reason to impose the Palatini variational principle in the standard Hilbert-Einstein theory instead of the metric variational principle.

However, the situation completely changes if we consider  ETGs, depending on functions of curvature invariants, as $f(R)$, or non-minimally coupled to some scalar field. In these cases, the Palatini and the metric variational principle provide different field equations and the theories thus derived differ \cite{magn, ferr}. The relevance of Palatini approach and the differences with respect to the metric approach, in this framework, have been recently proven in relation to cosmological applications \cite{f(R)-cosmo,voll}. 
From a physical viewpoint, considering the metric $g$ and the connection $\Gamma$ as independent fields means to decouple the metric structure of spacetime and its geodesic structure being, in general, the connection $\Gamma$ not the Levi-Civita connection of $g$ (see \cite{mauro} for details). The chronological
structure of spacetime is governed by $g$ while the trajectories of particles, moving in the spacetime, are governed by $\Gamma$.
This decoupling enriches the geometric structure of spacetime and generalizes the purely metric formalism. This metric-affine structure of spacetime is naturally translated, by means of the same (Palatini) field equations, into a bi-metric structure of spacetime. Beside the physical metric $g$, another metric $h$ is involved. This new metric is related, in the case of $f(R)$-gravity, to the connection. As a matter of fact, the connection $\Gamma$ results to be the Levi-Civita connection of $h$ and thus provides the geodesic structure of spacetime \cite{silvio}.

 In this paper, we consider the fact that a  large class of dynamical theories can be obtained in the form of a  {\it divergence theory} by defining currents $J^{\mu}$ in the dynamical mass term. In this context,  it is possible to demonstrate that the metric and Palatini actions become equivalent, provided we have  suitably defined massless scalar fields. This fact could be  extremely relevant in order to solve the ambiguities  that seem to emerge in the two approaches.

In Sec. (\ref{2}), we discuss the metric and Palatini formalisms for $f(R)$-gravity. In Sec. (\ref{3'}), we introduce a divergence free current theory and then, in this framework, we discuss equivalence of metric and Palatini formalisms. Conclusions are given in Sec. (\ref{4}).  

\section{Metric and Palatini formalisms for modified gravity}   \label{2}

The action  for $f(R)$-gravity takes the form
\begin{equation}\label{eq12'}
{\cal S}=\int_m d^4x \sqrt{-g}f({R})\,.
\end{equation}
In the metric formalism, the variation of the action is accomplished with respect to the metric. One can show that this action dynamically corresponds to an action of non-minimally coupled gravity with a new scalar field having no kinetic term.
By introducing a new auxiliary field $\chi$, the dynamically equivalent action is rewritten \cite{farasot, Sotiriou}
\begin{equation}\label{eq13'}
{\cal S}=\int_m d^4x
\sqrt{-g}(f(\chi)+f^\prime(\chi)(R-\chi)).
\end{equation}
Variation with respect to $\chi$ yields the equation
\begin{equation}\label{eq14'}
f^{\prime \prime}(\chi)({R}-\chi)=0\,.
\end{equation}
Therefore, $\chi=R$ if $f''(\chi)\neq0$ which reproduces the action (\ref{eq12'}).
Redefining the field $\chi$ by $\phi = f^\prime(\chi)$ and introducing  the potential
\beq
V(\phi)=\chi(\phi)\phi-f(\chi(\phi))\,,
\eneq
the action (\ref{eq13'}) takes the form
\begin{equation}\label{eq15'}
{\cal S}=\int_m d^4x \sqrt{-g}(\phi{R}-V(\phi)),
\end{equation}
that is the Jordan frame representation of the action of a Brans-Dicke theory with Brans-Dicke parameter $\omega_0=0$, known as O'Hanlon action in metric formalism.

Beside the metric formalism in which the variation of the action is done with respect to the metric, the Einstein equations can be derived as well
using the Palatini formalism, i.e., an independent variation with respect to the metric and an independent connection. The Riemann tensor and the Ricci tensor are also constructed with the independent connection and the metric is not needed to obtain the latter from the former. So, in order to
 stress the difference with respect to the metric formalism, we shall use ${\cal R}_{\mu\nu}$ and ${\cal R}$ instead of $R_{\mu \nu}$ and $R$, respectively. In the
ordinary Einstein-Hilbert action there is no specific difference between these two formalisms. However, once we generalize the action to depend on
a generalized form of the Ricci scalar they are no longer the same.

We briefly review the $f({\cal R})$-gravity in the
Palatini formalism and show that it corresponds to a Brans-Dicke
theory \cite{brans}. The action in the Palatini formalism without  matter is written
\begin{equation}\label{eq12}
{\cal S}=\int_p d^4x \sqrt{-g}f({\cal R}).
\end{equation}
By varying the action (\ref{eq12}) independently with respect to the
metric and the connection and using the formula
\begin{equation}\label{eq13}
\delta{\cal R}_{\mu \nu}=\bar{\nabla}_{\lambda}\delta
\Gamma^{\lambda}_{\mu \nu}-\bar{\nabla}_{\nu}\delta
\Gamma^{\lambda}_{\mu \lambda},
\end{equation}
yields
\begin{equation}\label{eq14}
f^{\prime}({\cal R}){\cal R}_{(\mu \nu)}-\frac{1}{2}f({\cal
R})g_{\mu \nu}=0,
\end{equation}
\begin{equation}\label{eq15}
\bar{\nabla}_{\lambda}(\sqrt{-g}f^{\prime}({\cal R})g^{\mu
\nu})-\bar{\nabla}_{\sigma}(\sqrt{-g}f^{\prime}({\cal R})g^{\sigma
(\mu })\delta_{\lambda}^{\nu)}=0,
\end{equation}
where $\bar{\nabla}$ denotes the covariant derivative defined with respect to
the independent connection $\Gamma^{\lambda}_{\mu \nu}$ and $(\mu
\nu)$ denotes the symmetry  over the indices $\mu, \nu$. Taking the
trace of Eq.(\ref{eq15}) gives rise to
\begin{equation}\label{eq16}
\bar{\nabla}_{\sigma}(\sqrt{-g}f^{\prime}({\cal R})g^{\sigma \mu
})=0,
\end{equation}
by which the field Eqs. (\ref{eq15}) becomes
\begin{equation}\label{eq17}
\bar{\nabla}_{\lambda}(\sqrt{-g}f^{\prime}({\cal R})g^{\mu
\nu})=0.
\end{equation}
Taking the trace of Eq.(\ref{eq14}) yields an algebraic equation in
${\cal R}$
\begin{equation}\label{eq18}
f^{\prime}({\cal R}){\cal R}-{2}f({\cal R})=0.
\end{equation}
 A  conformal metric to $g_{\mu \nu}$ can be defined as
\begin{equation}\label{eq19}
h_{\mu \nu}=f^\prime({\cal R})g_{\mu \nu},
\end{equation}
for which it is easily obtained that
\begin{equation}\label{eq20}
\sqrt{-h}h^{\mu \nu}=\sqrt{-g}f^\prime({\cal R})g^{\mu
\nu}.
\end{equation}
Eqs. (\ref{eq17}) are then the compatibility condition of the
metric $h_{\mu \nu}$ with the connection $\Gamma^{\lambda}_{\mu
\nu}$ and can be solved algebraically to give
\begin{equation}\label{eq21}
\Gamma^{\lambda}_{\mu \nu}=h^{\lambda \sigma}(\partial_{\mu} h_{\nu
\sigma}+\partial_{\nu} h_{\mu \sigma}-\partial_{\sigma} h_{\mu
\nu}).
\end{equation}
Under the conformal transformation (\ref{eq19}), the Ricci tensor and
its contracted form with $g^{\mu \nu}$ transform, respectively, as
\bea\label{eq22}
{\cal R}_{\mu \nu}=R_{\mu \nu}&+&\frac{3}{2}\frac{1}{(f^\prime({\cal
R}))^2}(\nabla_{\mu}f^\prime({\cal R}))(\nabla_{\nu}f^\prime({\cal
R}))+ \nonumber \\
&-&\frac{1}{(f^\prime({\cal R}))}(\nabla_{\mu}\nabla_{\nu}-\frac{1}{2}g_{\mu\nu}\Box)f^\prime({\cal R}),
\enea
\begin{equation}\label{eq23}
{\cal R}=R+\frac{3}{2}\frac{1}{(f^\prime({\cal
R}))^2}(\nabla_{\mu}f^\prime({\cal R}))(\nabla^{\mu}f^\prime({\cal
R}))-\frac{3}{(f^\prime({\cal R}))}\Box f^\prime({\cal
R}).
\end{equation}
Note the difference between ${\cal R}$ and the Ricci scalar of
$h_{\mu \nu}$ due to the fact that $g^{\mu \nu}$ is used here for
the contraction of ${\cal R}_{\mu \nu}$.
Now, by introducing a new auxiliary field $\chi$, the dynamically
equivalent action is rewritten \cite{farasot, Sotiriou}
\begin{equation}\label{eq24}
{\cal S}=\int_P d^4x
\sqrt{-g}(f(\chi)+f^\prime(\chi)({\cal R}-\chi)).
\end{equation}
Variation with respect to $\chi$ yields the equation
\begin{equation}\label{eq25}
f^{\prime \prime}(\chi)({\cal R}-\chi)=0.
\end{equation}
Redefining the field $\chi$ by $\phi = f^\prime(\chi)$ and
introducing  $V(\phi)=\chi(\phi)\phi-f(\chi(\phi))$ the action (\ref{eq24})
takes the form
\begin{equation}\label{eq26}
{\cal S}=\int_P d^4x \sqrt{-g}(\phi{\cal R}-V(\phi)).
\end{equation}
Now, we may use $\phi = f^\prime(\chi)$ in Eq.(\ref{eq23}) to write down
${\cal R}$ in terms of $R$ in the action (\ref{eq26}). This leads, modulus 
the surface term $\square \phi$ to
\begin{equation}\label{eq27}
{\cal S}=\int_P d^4x \sqrt{-g}\left(\phi
R+\frac{3}{2\phi}\nabla_{\mu}\phi \nabla^{\mu}\phi-V(\phi)\right).
\end{equation}
This is well known as the action in Palatini formalism which corresponds to a Brans-Dicke theory  with Brans-Dicke
parameter $\omega =-\frac{3}{2}$. Comparison of the action (\ref{eq15'})
with (\ref{eq27}) reveals that the former is the action of a Brans-Dicke
theory with $\omega_0=0$. Therefore, this is an indication that generally
the two actions are not dynamically equivalent.

\section{A divergence free current and the equivalence of Metric and Palatini formalisms}      \label{3'}

Let us now introduce a divergence theory developed in Ref. \cite{Sal}.
This theory starts by considering   a current. The original theory has been  developed
in the flat space-time, however it is easily generalized to the curved space-time. We have
\begin{equation}
J_{\mu}=-\frac{1}{2}\phi
\stackrel{\leftrightarrow}{\nabla_{\mu}}\phi^{-1}, \label{3}
\end{equation}
for which we obtain
\begin{equation}
\nabla_{\mu}J^{\mu}=\phi^{-1}[\Box \phi
-\phi^{-1}\nabla_{\mu}\phi \nabla^{\mu}\phi], \label{4}
\end{equation}
and
\begin{equation}
J_{\mu}J^{\mu}=\phi^{-2}\nabla_{\mu}\phi \nabla^{\mu}\phi,
\label{5}
\end{equation}
where $\phi$ is a real scalar field. Combining these relations
leads to
\begin{equation}
\Box \phi + \Gamma\{ \phi \} \phi=0, \label{6}
\end{equation}
where $\Gamma\{ \phi \}$ is called the {\it dynamical mass term}
\begin{equation}
\Gamma\{ \phi \}=-J_{\mu}J^{\mu}-\nabla_{\mu}J^{\mu}. \label{7}
\end{equation}
It is important to note that Eq.(\ref{6}) is a formal consequence
of the definition (\ref{3}), in the form of an identity, and it  is
not a dynamical equation for $\phi$. However, a large class of
dynamical theories may be obtained in the form of a {\it
divergence theory} by taking various currents $J^{\mu}$ in the
dynamical mass term. For example, a simple divergence theory is
developed by assuming
\begin{equation}
\nabla_{\mu}J^{\mu}=0, \label{8}
\end{equation}
which leads, through the field redefinition $\sigma=\ln \phi$, to
\begin{equation}
\Box \sigma=0. \label{9}
\end{equation}
Now, we may rewrite the action (\ref{eq26}) in which the surface term $\square \phi$, according to Eq.(\ref{eq23}), is being kept as follows
\begin{equation}\label{eq28}
{\cal S}=\int_P d^4x \sqrt{-g}\left(\phi
R+\frac{3}{2\phi}\nabla_{\mu}\phi \nabla^{\mu}\phi-3\square \phi-V(\phi)\right),
\end{equation}
or
\begin{equation}\label{eq29}
{\cal S}=\int_P d^4x \sqrt{-g}\left(\phi
R+3\phi\left[\frac{\nabla_{\mu}\phi \nabla^{\mu}\phi}{2\phi^2}-\frac{\square \phi}{\phi}\right]-V(\phi)\right).
\end{equation}
We now add and subtract the term ${\displaystyle \frac{3}{2\phi}\nabla_{\mu}\phi \nabla^{\mu}\phi}$ in the action which leads to
\bea   \label{eq30}
{\cal S}=\int_P d^4x \sqrt{-g}\left(\phi
R+3\phi\left[\frac{\nabla_{\mu}\phi \nabla^{\mu}\phi}{\phi^2}-\frac{\square \phi}{\phi}\right]+\right. \nonumber \\
-\left.\frac{3}{2\phi}\nabla_{\mu}\phi \nabla^{\mu}\phi-V(\phi)\right),
\enea
or
\begin{equation}\label{eq31}
{\cal S}=\int_P d^4x \sqrt{-g}\left(\phi
R-3\phi \nabla_{\mu}J^{\mu}-\frac{3}{2\phi}\nabla_{\mu}\phi \nabla^{\mu}\phi-V(\phi)\right).
\end{equation}
If we now demand the current $J^{\mu}$ to be divergence free,  we obtain \begin{equation}\label{eq32}
{\cal S}=\int_P d^4x \sqrt{-g}\left(\phi
R-\frac{3}{2\phi}\nabla_{\mu}\phi \nabla^{\mu}\phi-V(\phi)\right).
\end{equation}
Comparing the actions (\ref{eq32}) and (\ref{eq27}) reveals that the kinetic
term must be zero due to the sign difference. Therefore, the Palatini action reduces to the metric action
(\ref{eq15'}). This equivalence is a necessary result of a symmetry which
is included in the conservation equation $\nabla_{\mu}J^{\mu}=0$ with the
following current
\begin{equation}\label{eq33}
J_{\mu}=-\frac{1}{2}f^\prime(\chi)
\stackrel{\leftrightarrow}{\nabla_{\mu}}f^\prime(\chi)^{-1}.
\end{equation}
The conservation equation, by a field redefinition 
\beq
\sigma=\ln f^\prime(\chi),   \label{eq34}
\eneq 
may be encoded in a massless scalar field equation
\begin{equation}\label{eq35}
\Box \sigma=0.
\end{equation}
Therefore, the metric and Palatini actions become equivalent provided we have the massless scalar field $\sigma=\ln f^\prime(\chi)$. 
 In the case of Einstein's General Relativity, where $f(\chi)=\chi$, the scalar field equation is the trivial identity $0=0$, which indicates the equivalence between metric and Palatini formalisms.

\section{General Remarks}            \label{4}

In the context of $f(R)$-gravity, we have introduced a dynamically equivalent action both in the metric frame (the O'Hanlon action in metric formalism (\ref{eq15'})) and in the Palatini formalism (\ref{eq26}). 
 In the framework of Palatini formalism,  we have used a  conformal transformation of the metric $g_{\mu\nu}$ in Eq. (\ref{eq19}) and expressed the conformally transformed Ricci scalar $R$ in Eq. (\ref{eq23}) by the redefinition $\phi = f^\prime(\chi)$. Then we  have demonstrated the equivalence between metric and Palatini formalisms in $f(R)$ gravity in the general context of theories with divergence free current. In this way,  we  focused that the two  approaches can be directly related by conformal transformations in a very straightforward way. 
In fact, considering a simple divergence theory, and the current suitably defined in terms of the field $\phi$, it is possible to generalize, as we have demonstrated, the conformal equivalence between them. This equivalence is a result which can be included in the conservation equation of current which manifests an important symmetry. In fact, through an appropriate redefinition of the dynamical auxiliary field that we have introduced in the context of dynamically equivalent action, the conservation equation may be encoded in a massless scalar field equation. 
 In this way, we have obtained that the metric and Palatini formalisms become dinamically equivalent if there is the presence of the massless scalar field suitably defined by Eq. (\ref{eq34}). 

\section*{Acknowledgment}
This work has been supported by ``Research Institute for
Astronomy and Astrophysics of Maragha (RIAAM)'', Iran. F. Darabi would like to appreciate R.I.A.A.M for this hospitality.

\end{document}